\documentclass{article}
\usepackage{spconf,amsmath,graphicx,hyperref}

\usepackage[table]{xcolor}
\usepackage{tabularx}
\usepackage{booktabs}
\usepackage{amssymb} 
\usepackage{amsfonts}
\usepackage{arydshln}
\usepackage{setspace}   
\usepackage[numbers,sort&compress]{natbib}  

\usepackage{array} 

\makeatletter
\renewcommand\subsection{\@startsection{subsection}{2}{\z@}%
  {0.65ex plus .2ex minus .2ex}
  {0.35ex}
  {\normalfont\normalsize\bfseries}}

\makeatother

\title{LAMB: LLM-based Audio Captioning with Modality Gap Bridging \\via Cauchy-Schwarz Divergence}
\name{Hyeongkeun Lee$^{*}$, Jongmin Choi$^{*}$, KiHyun Nam, Joon Son Chung\thanks{$^*$These authors contributed equally to this work.}}
\address{Korea Advanced Institute of Science and Technology, South Korea
\\\{lhk528, cjmin, nkh.mmai, joonson\}@kaist.ac.kr}
\newcommand{\newpara}[1]{\vspace{2pt}\noindent\textbf{#1}}

\newcommand{\compressmath}{%
  \setlength\abovedisplayskip{4pt}%
  \setlength\belowdisplayskip{4pt}%
  \setlength\abovedisplayshortskip{2pt}%
  \setlength\belowdisplayshortskip{2pt}%
}

\begin{document}
%
\maketitle
\begin{abstract}
Automated Audio Captioning (AAC) aims to describe the semantic content of input audio. Recent works have employed large language models (LLMs) as a text decoder to leverage their reasoning capabilities. However, prior approaches that project audio features into the LLM embedding space without considering cross-modal alignment fail to fully utilize these capabilities. To address this, we propose \textit{LAMB}, an LLM-based audio captioning framework that bridges the modality gap between audio embeddings and the LLM text embedding space. LAMB incorporates a Cross-Modal Aligner that minimizes Cauchy–Schwarz (CS) divergence while maximizing mutual information, yielding tighter alignment between audio and text at both global and token levels. We further design a Two-Stream Adapter that extracts semantically enriched audio embeddings, thereby delivering richer information to the Cross-Modal Aligner. Finally, leveraging the aligned audio embeddings, a proposed Token Guide directly computes scores within the LLM text embedding space to steer the output logits of generated captions. Experimental results confirm that our framework strengthens the reasoning capabilities of the LLM decoder, achieving state-of-the-art performance on AudioCaps. Code is available at \url{https://github.com/Hyeongkeun/LAMB}.

\end{abstract}
\section{Introduction}
Automated Audio Captioning (AAC) is a multimodal task that aims to generate corresponding descriptions for given audio content~\cite{xu2023beyond}. Capturing acoustic semantic information and aligning it with textual features are key challenges in AAC, as these processes enable the model to generate semantically consistent and contextually appropriate descriptions.

Prior works~\cite{kim2023prefix,tang2023salmonn,kim2024avcap,rho2025lavcap,liu2024enhancing,chen2025slam,kim2024enclap,haji2024taming,takeuchi2025clap,choi2025enhancing} typically follow a two-stage framework, using an audio encoder~\cite{kong2020panns,chen2023beats,elizalde2023clap} and a text decoder~\cite{devlin2019bert,radford2019language,lewis2020bart}.
More recently, with the emergence of large language models (LLMs), studies~\cite{tang2023salmonn,liu2024enhancing,kim2024avcap,rho2025lavcap,chen2025slam, nam2025diffusion} have employed LLMs~\cite{chiang2023vicuna,touvron2023llama,chung2024scaling} as a text decoder to leverage their contextual reasoning.

However, because LLMs are trained on large-scale text corpora, their reasoning is most effectively exploited when the conditioning inputs (e.g., audio modality) are represented in the LLM text embedding space~\cite{masry2025alignvlm}. Existing methods~\cite{tang2023salmonn,liu2024enhancing,chen2025slam} map audio features into this space, typically via linear projection or Q-Former~\cite{li2023blip}, without an explicit objective that bridges the audio–text modality gap. As a result, a modality misalignment can remain between audio representations and the LLM embedding space, limiting the ability of the decoder that interprets audio semantics.

To overcome this limitation, we introduce \textbf{\textit{LAMB}}, an \textbf{L}LM-based \textbf{A}udio Captioning framework for \textbf{M}odality \textbf{B}ridging.
Our approach lies in a \textit{Cross-Modal Aligner} that leverages Cauchy-Schwarz (CS) divergence~\cite{principe2000learning}. This symmetric metric, combined with mutual information, provides a robust distance estimation that effectively captures both global structures and pairwise semantic correspondences between different modalities~\cite{yin2025distributional}.
To the best of our knowledge, this is the first use of CS divergence to mitigate the modality gap for LLM-driven audio captioning.
Ensuring effective cross-modal alignment, the audio features are processed with \textit{Two-Stream Adapter}, which extracts both semantically and temporally rich context. With these high-quality audio features, the \textit{Cross-Modal Aligner} can more accurately align them to the text embedding space. Finally, our novel \textit{Token Guide} directly steers the text generation process. It handles scores between the aligned audio and token embeddings using only the LLM's own token dictionary, avoiding reliance on auxiliary modules that constrained previous works~\cite{chen2025slam}.

Experiments validate the effectiveness of each component of the proposed method. Furthermore, our framework achieves superior performance over state-of-the-art methods on AudioCaps, and shows comparable results with some improvements on Clotho.

\begin{figure*}[t]
\centering
\includegraphics[width=0.95\textwidth]{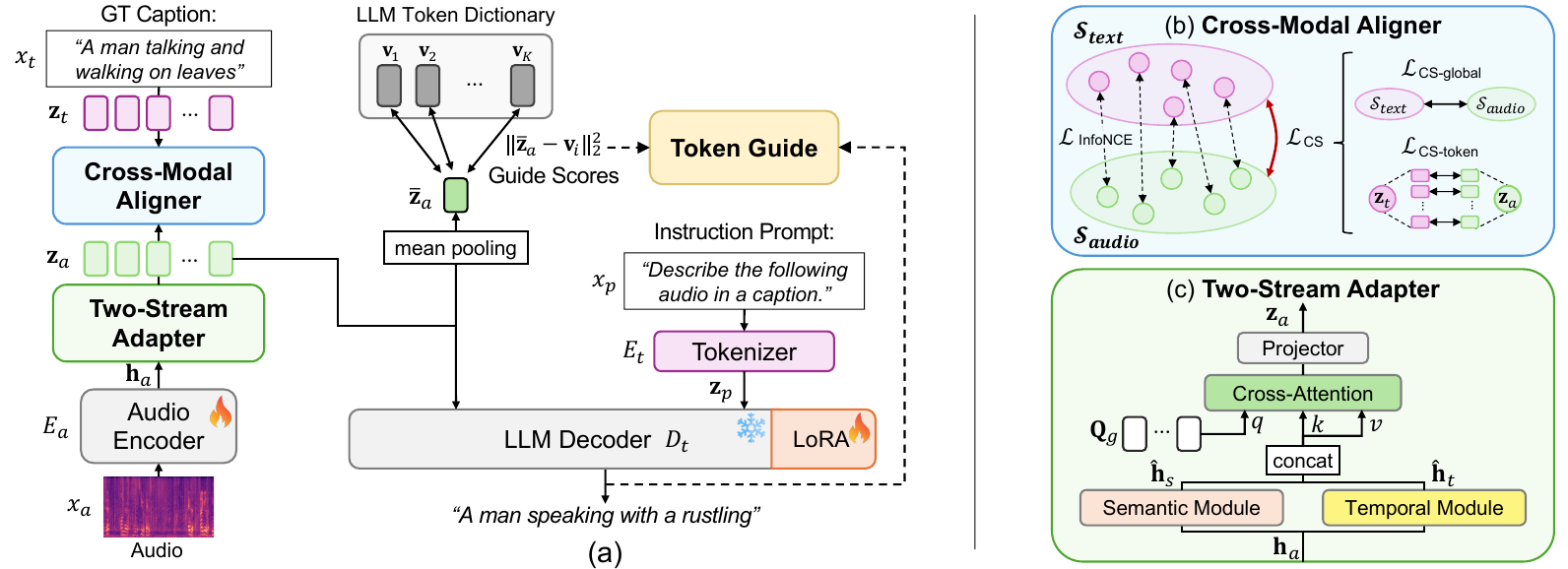}
\vspace{-0.8em}
\caption{(a) Overview of the proposed LAMB framework with (b) Cross-Modal Aligner and (c) Two-Stream Adapter.}
\label{fig:main_fig}
\vspace{-1em}
\end{figure*}
\vspace{-0.7em}
\section{Method}
Fig.~\ref{fig:main_fig} shows the overall architecture of LAMB, and the subsequent sections provide detailed explanations of each component and the training objectives.

\subsection{Two-Stream Adapter}
The input audio $x_a$ is first encoded by the audio encoder $E_a$ into embeddings $\mathbf{h}_a = E_a(x_a) \in \mathbb{R}^{N_a \times D}$, where $N_a$ is the number of audio tokens and $D$ the embedding dimension. We then introduce the \textit{Two-Stream Adapter} to capture semantic and temporal context of $\mathbf{h}_a$ through two parallel modules.

\newpara{Semantic Module.}
Semantic features are extracted by $N_s$ learnable queries $\mathbf{Q}_s \in \mathbb{R}^{N_s \times D}$ attending to $\mathbf{h}_a$ through multi-head attention (MHA), followed by residual connection and LayerNorm (LN):
{\compressmath
\begin{equation}
\scalebox{0.9}{$\mathrm{MHA}(\mathbf{Q}_s,\mathbf{h}_a)=\sigma\left(\frac{\mathbf{Q}_sW_Q\cdot (\mathbf{h}_aW_K)^\top}{\sqrt{d_H}}\right)\mathbf{h}_aW_V,$}
\end{equation}
\begin{equation}
\scalebox{0.9}{$\hat{\mathbf{h}}_s=\mathrm{LN}(\mathrm{MHA}(\mathbf{Q}_s,\mathbf{h}_a)+\mathbf{Q}_s),$}
\end{equation}}
where $\sigma$ is softmax, $W_Q,W_K,W_V \in \mathbb{R}^{D \times Hd_H}$ are projection matrices, $H$ the number of heads, and $d_H$ the head size.
\newpara{Temporal Module.}
Temporal features are extracted using a multi-scale 1D convolutional front-end, a two-layer bidirectional GRU~\cite{chung2014empirical}, and MHA with residual connection and LayerNorm. The convolutional front-end captures local patterns from $\mathbf{h}_a$, which are projected to dimension $D$ with LayerNorm and GELU activation. The GRU encodes contextual dependencies, and $N_t$ learnable queries $\mathbf{Q}_t \in \mathbb{R}^{N_t \times D}$ attend to its outputs to form temporal representations:
{\compressmath
\begin{equation}
\scalebox{0.9}{$\displaystyle \hat{\mathbf{h}}_t = \mathrm{LN}(\mathrm{MHA}(\mathbf{Q}_t,\mathrm{GRU}(\mathrm{Conv1D}(\mathbf{h}_a)))+\mathbf{Q}_t),$}
\end{equation}}
where $\mathrm{Conv1D}(\cdot)$ and $\mathrm{GRU}(\cdot)$ denote the convolutional front-end and GRU encoder.

\newpara{Fusion.} The outputs of the two parallel modules $\hat{\mathbf{h}}_s$ and $\hat{\mathbf{h}}_t$ are concatenated along the token axis, and then processed with $N_g$ global queries $\mathbf{Q}_g$ via a cross-attention layer with residual and LayerNorm. The resulting representations are projected into the $D_{llm}$-dimensional LLM text embedding space via $W_p \in \mathbb{R}^{D \times D_{llm}}$, yielding $\mathbf{z}_a\in\mathbb{R}^{N_g\times D_{llm}}$:
{\compressmath
\begin{equation}
\scalebox{0.9}{$\displaystyle \mathbf{z}_a = \left(\mathrm{LN}(\mathrm{MHA}(\mathbf{Q}_g,[\hat{\mathbf{h}}_{s};\hat{\mathbf{h}}_{t}])+\mathbf{Q}_g)\right)W_p.$}
\end{equation}}
\begin{table*}[t!]
    \caption{AAC results on AudioCaps and Clotho. Bold numbers denote the best scores and underlined numbers the second best. Gray-shaded rows indicate results with additional pretraining datasets. $^*$Reported from previous work~\cite{chen2025slam}. Pre-training datasets include AudioCaps (AC), Clotho (CL), augmented Clotho (CL$_P$~\cite{chen2025slam}), WavCaps (WC), MACS (MA), and AutoACD (AA). Evaluation metrics are METEOR (MT), CIDEr (CD), SPICE (SC), SPIDEr (SD), SPIDEr-FL (SF), and FENSE (FS).}
    \centering
    \small
    \begin{tabularx}{\textwidth}{lc>{\centering\arraybackslash}X>{\centering\arraybackslash}X>{\centering\arraybackslash}X>{\centering\arraybackslash}X>{\centering\arraybackslash}X>{\centering\arraybackslash}X>{\centering\arraybackslash}X>{\centering\arraybackslash}X>{\centering\arraybackslash}X>{\centering\arraybackslash}X>{\centering\arraybackslash}X>{\centering\arraybackslash}X}
        \toprule
        & & \multicolumn{6}{c}{AudioCaps}&\multicolumn{6}{c}{Clotho}\\
        \cmidrule(lr){3-8} \cmidrule(lr){9-14}
        Model & PT Dataset  & MT& CD& SC& SD& SF& FS& MT& CD& SC& SD& SF& FS\\
        \midrule
        HTSAT-BART~\cite{mei2024wavcaps}  & AC+CL+WC & 25.0 & 78.7 & 18.2 & 48.5 & 48.3$^*$ & 64.2$^*$  & 18.5 & 48.8 & 13.3 & 31.0 & 29.6$^*$ & 50.1$^*$ \\
        EnCLAP\small{-large}~\cite{kim2024enclap} & AC+CL & 25.5 & 80.3 & 18.8 & 49.5 & 49.9$^*$ & 65.5$^*$ & 18.6 & 46.4 & 13.3 & 29.9 & 28.9$^*$ & 50.7$^*$ \\
        AutoCap~\cite{haji2024taming} & AC+CL+WC  & 25.6 & 80.4 & 19.0 &  49.7 & -& -& -& -& -& - & -& -  \\
        LOAE~\cite{liu2024enhancing} & AC+CL+WC  & 26.7 & 81.6 & 19.3 & 50.5 & 50.4 & 66.4  & \underline{19.7} & 51.3 & \underline{14.7} & 33.0 & \underline{33.0} & \underline{53.8}    \\
        CLAP-ART~\cite{takeuchi2025clap} & - / AC & 25.6 & 80.7 & 18.8 & 49.8 & -& 65.5 & 18.7 & 47.5 & 13.3 &  30.4 & - & 51.1     \\
        \rowcolor{gray!10}
        MQ-Cap~\cite{choi2025enhancing} & AC+CL+WC+AA & - & \underline{84.5} & \underline{19.4} & \underline{51.9} & - & - & - & 49.6 & 14.3 & 31.9 & - & -     \\
        \rowcolor{gray!10}
        SLAM-AAC~\cite{chen2025slam} & AC+CL$_P$ +WC +MA & \underline{26.8} & 84.1 & \underline{19.4} &  51.8 & \underline{51.5} & \underline{66.8} & \underline{19.7} & \underline{51.5} & \textbf{14.8} &  \underline{33.2} & \underline{33.0} & \textbf{54.0} \\    
        \hdashline
        LAMB (ours) & AC+CL+WC & \textbf{27.1} & \textbf{91.1} & \textbf{19.7} & \textbf{55.4} & \textbf{55.3} & \textbf{67.7} & \textbf{19.8} & \textbf{52.3} & \underline{14.7} & \textbf{33.4} & \textbf{33.1} & 53.4\\     
        \bottomrule
    \label{tab:main_tab}
  \end{tabularx}
  \vspace{-2.5em}
\end{table*}
\subsection{Cross-Modal Aligner}

A \textit{Cross-Modal Aligner} employs Cauchy–Schwarz (CS) divergence with mutual information to align audio embeddings $\mathbf{z}_a$ with caption embeddings $\mathbf{z}_t$ in a batch $B$, where $\mathbf{z}_t$ is the encoding of caption $x_t$ with the LLM tokenizer $E_t$. 
The representative vectors $\bar{\mathbf{z}}_a$ and $\bar{\mathbf{z}}_t$ are computed via mean pooling to quantify the modality gap between audio and the LLM text distributions. The CS divergence, $D_\text{CS}$, provides a symmetric and robust measure of the distance between the distributions of audio $p_a$ and text $p_t$ as follows:
{\compressmath
\begin{equation}
\scalebox{0.9}{$\displaystyle D_\text{CS}(p_a; p_t)
= -\log \frac{\int p_a(\omega) p_t(\omega)\, d\omega}
{\sqrt{\int p_a^2(\omega)\, d\omega \int p_t^2(\omega)\, d\omega}},$}
\label{eq:cs_div}
\end{equation}
}
with $0 \leq D_\text{CS} < \infty$ and $D_\text{CS}=0$ if and only if $p_a=p_t$. 

\newpara{Global-level CS Divergence.}
Since Eq.~\ref{eq:cs_div} is intractable, we approximate $D_\text{CS}$ via non-parametric KDE\cite{parzen1962estimation} using i.i.d. samples $\{\bar{\mathbf{z}}_a^{(i)}\}_{i=1}^B \sim p_a(\bar{\mathbf{z}}_a)$ and $\{\bar{\mathbf{z}}_t^{(i)}\}_{i=1}^B \sim p_t(\bar{\mathbf{z}}_t)$, where each batch $B$ spans the entire dataset, to measure the \textit{global-level} distributional distance, as follows:
{\compressmath
\begin{equation}
\scalebox{0.86}{$\displaystyle \widehat{D}_{\text{CS}}(p_a(\bar{\mathbf{z}}_a);p_t(\bar{\mathbf{z}}_t))
= -\log \frac{\sum_{i,j=1}^B \kappa_{at}^{ij}}
{\sqrt{\Big(\sum_{i,j=1}^B \kappa_{aa}^{ij}\Big)
       \Big(\sum_{i,j=1}^B \kappa_{tt}^{ij}\Big)}},$}
\end{equation}}
where $\kappa^{ij}_{mn}=\kappa(\bar{\mathbf{z}}^{(i)}_m,\bar{\mathbf{z}}^{(j)}_n)$, $\kappa$ is Gaussian kernel $\kappa_\sigma(\textbf{x},\textbf{y})=\exp(-||\textbf{x}-\textbf{y}||_2^2/2\sigma^2)$ with kernel width $\sigma$. 
The global distribution distance is reduced by the loss $\mathcal{L}_\text{CS-global}=\widehat{D}_\text{CS}$.

\newpara{Token-level CS Divergence.}
To achieve fine-grained alignment between audio and caption embeddings, we present a token-level loss based on token-wise CS divergence. For each pair of token-level embeddings $\{\mathbf{z}^{(i)}_a,\mathbf{z}^{(i)}_t\}^B_{i=1}$, the CS divergence is estimated as $\widehat{D}_\text{CS}(p_a(\mathbf{z}^{(i)}_a);p_t(\mathbf{z}^{(i)}_t))$ and the token-level loss is defined as the average of per-sample divergences:
{\compressmath
\begin{equation}
\scalebox{0.9}{$\displaystyle 
    \mathcal{L}_\text{CS-token}=\frac{1}{B}\sum^B_{i=1}\widehat{D}_\text{CS}(p_a(\mathbf{z}^{(i)}_a);p_t(\mathbf{z}^{(i)}_t)).$}
\end{equation}
}

\begin{table}[t!]
\caption{Ablation of alignment methods on AudioCaps.}
\centering
\small
\begin{tabularx}{\columnwidth}{l>{\centering\arraybackslash}X>{\centering\arraybackslash}X>{\centering\arraybackslash}X>{\centering\arraybackslash}X>{\centering\arraybackslash}X>{\centering\arraybackslash}X}
\toprule
Align Method & $L_2$ $\downarrow$ & Cos$\uparrow$ & MT$\uparrow$ & CD$\uparrow$ & SC$\uparrow$ & SD$\uparrow$\\
\midrule
Q-former & 111.8 & 0.04 & 24.4  & 66.9  & 18.1  & 42.5 \\
Linear Layer & 62.6  & -.01 & 25.3  & 77.2  & 18.1  & 47.7 \\
CMA (ours) & \textbf{10.9} & \textbf{0.58}  & \textbf{27.1}      & \textbf{91.1}     & \textbf{19.7}       & \textbf{55.4}          \\
\bottomrule
\end{tabularx}
\label{tab:abl_gap}
\vspace{-1em}
\end{table}
\begin{figure}[t]
\centering
\includegraphics[width=\columnwidth]{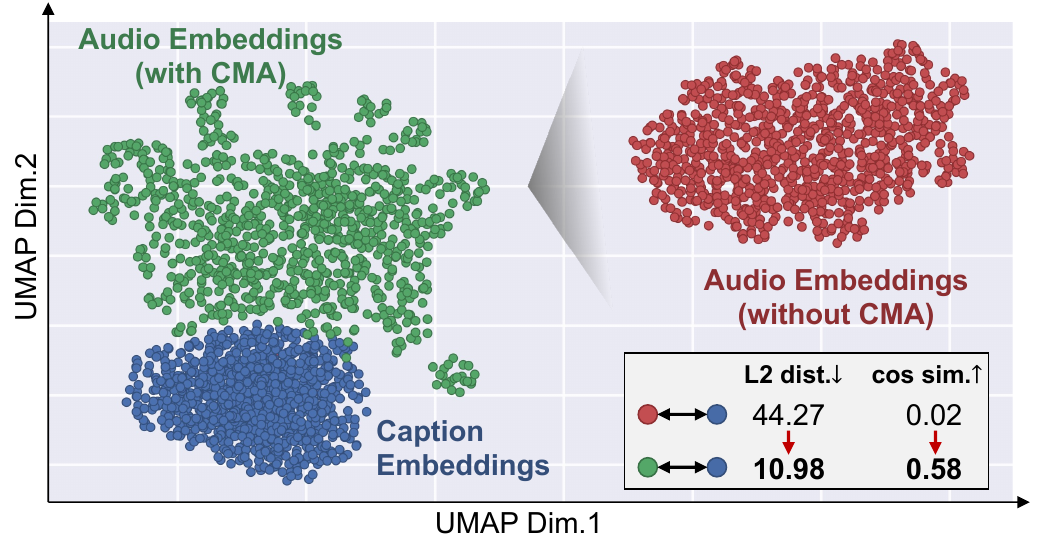}
\vspace{-2.2em}
\caption{UMAP Visualization of audio–text embedding alignment with/without Cross-Modal Aligner (CMA).}
\label{fig:exp_gap}
\vspace{-1.4em}
\end{figure}

Lastly, we adopt the InfoNCE loss~\cite{oord2018representation} to maximize lower bounds of mutual information between modalities, defined as:
{\compressmath
\begin{equation}
\scalebox{0.9}{$\displaystyle 
\mathcal{L}_{\text{InfoNCE}}
=\tfrac{1}{2}\big(\mathcal{L}_{a\rightarrow t}+\mathcal{L}_{t\rightarrow a}\big),$}
\end{equation}
\begin{equation}
\scalebox{0.9}{$\mathcal{L}_{m\rightarrow n}
= -\tfrac{1}{B}\sum_{i=1}^B
\log \frac{\exp\big(\mathrm{sim}(\bar{\mathbf{z}}_m^i,\bar{\mathbf{z}}_n^i)/\tau\big)}
{\sum_{j=1}^B \exp\big(\mathrm{sim}(\bar{\mathbf{z}}_m^i,\bar{\mathbf{z}}_n^j)/\tau\big)}.$}
\end{equation}}
Here, $\tau$ is the temperature and $\mathrm{sim}(\cdot,\cdot)$ the cosine similarity.

The overall cross-modal alignment loss is as follows:
{\compressmath
\begin{equation}
\scalebox{0.95}{$\displaystyle \mathcal{L}_{\text{cma}}
= \alpha_1 \mathcal{L}_{\text{CS-global}}
+ \alpha_2 \mathcal{L}_{\text{CS-token}}
+ \alpha_3 \mathcal{L}_{\text{InfoNCE}}.$}
\end{equation}
}

\subsection{Text Decoding with Token Guide}
We propose \textit{Token Guide} to steer the output logits using solely the LLM token dictionary without auxiliary module or external corpora. At each decoding step $l$, the logits $\mathbf{z}_l$ of LLM decoder $D_t$ are obtained from the concatenation of the audio embedding $\mathbf{z}_a$ and the instruction prompt embedding $\mathbf{z}_p=E_t(x_p)$, where $E_t$ is the LLM tokenizer:
{\compressmath
\begin{equation}
    \scalebox{0.95}{$\displaystyle \mathbf{z}_l=D_t([\mathbf{z}_a;E_t(x_p)],M_{att}),$}
\end{equation}}
where $M_{att}$ is the attention mask enforcing the autoregressive property of the LLM. At each step $l$, the guided logit $\hat{z}_{l,i}$ is obtained by correcting the original logit $z_{l,i}$ with \textit{Guide Scores}, squared $\mathrm{L_2}$ distance between global mean $\bar{z}_a$ and token embeddings $\{\mathbf{v}_i\}^K_{i=1}$, scaled by a learnable coefficient $\beta$:
{\compressmath
\begin{equation}
    \scalebox{0.9}{$\displaystyle \hat{z}_{l,i}=z_{l,i}-\beta\cdot||\bar{\mathbf{z}}_a-\mathbf{v}_i||^2_2,\quad i=1,\dots,K. $}
\end{equation}}
Given the audio embeddings $\mathbf{z}_a$ and prompt embeddings $\mathbf{z}_p$, 
the autoregressive cross-entropy loss is defined as
{\compressmath
\begin{equation}
\scalebox{0.95}{$\displaystyle \mathcal{L}_{\text{ce}}
= -\frac{1}{L}\sum_{l=1}^L \log \big[p(y_l|y_{<l}, \mathbf{z}_a, \mathbf{z}_p)\big],$}
\end{equation}}
where $p(\cdot)$ denotes the predicted probability distribution at step $l$. 
Both the original logits ($p_{\text{orig}}=\mathrm{Softmax}(\mathbf{z}_l)$) and the guided logits ($p_{\text{tg}}=\mathrm{Softmax}(\hat{\mathbf{z}}_l)$) are utilized to compute the loss functions $\mathcal{L}_{\text{dec}}$ and $\mathcal{L}_{\text{tg}}$, respectively.

\subsection{Loss Functions}
The final objective combines all components with weights:
{\compressmath
\begin{equation}
\scalebox{0.95}{$\displaystyle \mathcal{L}_{\text{total}}
\;=\; \lambda_1 \mathcal{L}_{\text{cma}}
+ \lambda_2 \mathcal{L}_{\text{tg}}
+ \lambda_3 \mathcal{L}_{\text{dec}}.$}
\end{equation}}

\section{Experiments}

\subsection{Experimental Settings}
\subsubsection{Datasets}
We pre-train the model on the combined dataset of AudioCaps~\cite{kim2019audiocaps}, Clotho~\cite{drossos2020clotho}, and WavCaps~\cite{mei2024wavcaps}, and fine-tune it separately on AudioCaps and Clotho. AudioCaps provides 48,595 training and 944 test clips (10 s each) with audio-based annotations, retrieved from available links. Clotho v2.1 contains 3,839 development, 1,045 validation, and 1,045 evaluation clips (15–30 s), each with five captions. WavCaps comprises 403,050 clips (a few seconds to over one minute) collected from AudioSet~\cite{gemmeke2017audio}, BBC Sound Effects,\footnote{https://sound-effects.bbcrewind.co.uk} FreeSound,\footnote{https://freesound.org} and SoundBible.\footnote{https://soundbible.com}

\subsubsection{Implementation Details and Metrics}
We adopt the consistent ensemble distillation~\cite{dinkel2024ced} model as the audio encoder and LLaMA 2 (7B)~\cite{touvron2023llama} as the text decoder, fine-tuned with LoRA~\cite{hu2022lora}. Query numbers are set to $N_s$=8, $N_t$=8, and $N_g$=32.
All experiments used AdamW with weight decay 1e-6, trained for 30 epochs (2 warmup epochs) under a cosine annealing schedule. Pre-training employed a learning rate of 5e-5 with batch size 32, and fine-tuning used 3e-6 with batch size 8.
For evaluation, we adopt common AAC metrics, including METEOR~\cite{banerjee2005meteor}, CIDEr~\cite{vedantam2015cider}, SPICE~\cite{anderson2016spice}, SPIDEr~\cite{liu2017improved}, SPIDEr-FL~\cite{zhou2022can}, and FENSE~\cite{zhou2022can}.

\begin{table}[t!]
\caption{Ablation of Guide Scores on AudioCaps.}
\centering
\small
\begin{tabularx}{\columnwidth}{l>{\centering\arraybackslash}X>{\centering\arraybackslash}X>{\centering\arraybackslash}X>{\centering\arraybackslash}X>{\centering\arraybackslash}X>{\centering\arraybackslash}X}
\toprule
Guide Scores  & MT & CD & SC & SD & SF & FS \\
\midrule
Cosine sim.  & 26.6      & 81.6      & 19.0      & 50.3      & 50.2 & 66.8 \\
$L_1$ dist.  & 25.9      & 88.6      & 17.5      & 53.1      & 52.4 & 65.5 \\
$L_2$ dist. (ours)  & \textbf{27.1}      & \textbf{91.1}     & \textbf{19.7}       & \textbf{55.4} & \textbf{55.3}    & \textbf{67.7}    \\

\bottomrule
\end{tabularx}
\label{tab:abl_guide}
\vspace{-1em}
\end{table}

\begin{table}[t!]
\caption{Ablation of LAMB components of on AudioCaps.}
\centering
\small
\begin{tabularx}{\columnwidth}{l>{\centering\arraybackslash}X>{\centering\arraybackslash}X>{\centering\arraybackslash}X>{\centering\arraybackslash}X>{\centering\arraybackslash}X>{\centering\arraybackslash}X}
\toprule
Element Ablation & MT & CD & SC & SD & SF & FS \\
\midrule
LAMB (ours)     & \textbf{27.1}      & \textbf{91.1}     & \textbf{19.7}       & \textbf{55.4} & \textbf{55.3}    & \textbf{67.7}     \\
\; w/o TSA   & 26.7      & 87.6      & 18.9      & 53.3      & 53.2      & 65.9 \\
\; w/o $\mathcal{L}_\text{CS}$ ($\mathcal{L}_\text{CS-global+CS-token}$)  & 24.4      & 84.8      & 16.6      & 50.7      & 49.4      & 63.0 \\
\; w/o CMA ($\mathcal{L}_\text{cma}$)   & 25.8      & 78.6      & 18.9      & 48.7      & 48.6      & 65.8 \\
\; w/o TG & 26.8      & 82.2      & 19.2      & 50.7      & 50.6      & 66.9 \\
\bottomrule
\end{tabularx}
\label{tab:abl_comp}
\vspace{-1em}
\end{table}

\subsection{Main Results}
Tab.~\ref{tab:main_tab} summarizes the performance of LAMB on the AAC benchmarks. LAMB surpasses prior work on \emph{all} evaluation metrics, establishing clear state-of-the-art results on AudioCaps. On AudioCaps, it achieves notable gains, particularly on CIDEr, SPIDEr, and SPIDEr-FL, while also improving METEOR and SPICE, reflecting stronger semantic and structural fidelity. 
On Clotho, which is more challenging due to diverse captions, LAMB shows competitive performance with improvements on several metrics, demonstrating robustness to annotation variability. Notably, while several prior works rely on additional pre-training data, our model attains comparable or better results across the datasets.

These results show that bridging the modality gap enhances the LLM’s reasoning and produces captions that are more faithful, coherent, and aligned with human preference, reflected in improvements in FENSE and SPIDEr-FL.

\subsection{Ablation Studies}
\newpara{Modality Gap Alignment Strategies.}
We first assess the effectiveness of the Cross-Modal Aligner (CMA). As shown in Tab.~\ref{tab:abl_gap}, CMA yields the lowest $L_2$ distance and highest cosine similarity, surpassing linear and Q-Former baselines. Moreover, Fig.~\ref{fig:exp_gap} visualizes with UMAP that CMA clearly reduces the modality gap compared to the case without CMA.

\newpara{Token Guide Metrics.}
Tab.~\ref{tab:abl_guide} presents the ablation results on identifying effective methods for computing Guide Scores in Token Guide (TG).
Squared $L_2$ distance proves most effective, stably steering decoder logits and achieving the highest scores, while $L_1$ distance slightly degrades performance and cosine similarity yields results inferior to the baseline without guidance.
Even without auxiliary modules or external knowledge, Guide Scores derived from the LLM text embedding space effectively guide the decoder.

\newpara{Effect of Components.}
Tab.~\ref{tab:abl_comp} highlights the role of each module by ablating components of LAMB. Two-Stream Adapter (TSA) provides semantic and temporal features of audio to CMA, and its removal reduces both semantic coverage and structural fidelity. Within CMA, two loss terms are used: CS divergence and InfoNCE. Removing $\mathcal{L}_\text{CS}$ causes a larger performance drop (rows 1 $\&$ 3) than removing $\mathcal{L}_\text{InfoNCE}$ (rows 3 $\&$ 4), and excluding both further lowers all metrics. TG is also important, as its absence lowers performance across all metrics. Overall, these modules offer complementary benefits, with the full model achieving the best results.

\label{sec/abl}

\subsection{Qualitative Results}
In Tab.~\ref{tab:qual_cap_token_guide}, we present captions generated by LAMB trained with and without TG. Since the Guide Scores are computed as the squared $L_2$ distance between aligned audio embeddings and LLM token embeddings, semantically related tokens receive higher weights, enabling more accurate and precise guidance. In the first example, TG allows the model to capture not only the sound of a basketball bouncing but also people talking, generating details such as \textit{``hard surface"} and \textit{``a group of people talk in the background"}. In the second example, TG links \textit{``loud crash"} to \textit{``glass shattering"} and additionally identifies background sounds such as \textit{``a rooster crows"} and \textit{``birds chirping"}. These results indicate that TG can directly guide in the LLM text embedding space, enabling precise and accurate caption generation.

\begin{table}[t!]
  \renewcommand{\arraystretch}{1}
  \footnotesize
  \caption{Qualitative results without and with Token Guide. Underlined text indicates audio cues captured more explicitly.}
  \centering
  \newcommand{\audio}[1]{\textcolor{red}{\textbf{#1}}}
  \begin{tabularx}{\linewidth}{XX}
    \toprule
    \multicolumn{1}{c}{w/o Token Guide} & \multicolumn{1}{c}{w Token Guide}\\
    \midrule
    \begin{minipage}{\hsize} \emph{“Basketballs are dribbled and shoes squeak as a man speaks”} \end{minipage} &
    \begin{minipage}{\hsize} \emph{“Sneakers squeaking and basketballs bouncing on \underline{\textbf{a hard surface}} as \underline{\textbf{a group of people talk in the}} \underline{\textbf{background}}”} \end{minipage} \\
    \midrule
    \begin{minipage}{\hsize}\emph{“A loud crash followed by a man speaking and a woman screaming”} \end{minipage} &
    \begin{minipage}{\hsize} \emph{“\underline{\textbf{Glass shattering}} followed by a man speaking then a woman speaking before \underline{\textbf{a rooster crows}} and \underline{\textbf{birds chirping}}”} \end{minipage} \\
    \bottomrule
  \end{tabularx}
\label{tab:qual_cap_token_guide}
\vspace{-1.5em}
\end{table}
\section{Conclusion}
In this work, we introduced \textbf{LAMB}, an LLM-based audio captioning framework that, for the first time, applies Cauchy-Schwarz divergence to bridge the audio-text modality gap in audio captioning.
LAMB integrates a Cross-Modal Aligner for fine-grained alignment and a Two-Stream Adapter to enrich audio representations, while the Token Guide steers decoder logits using Guide Scores in the LLM text embedding space.
Experimental results show that LAMB outperforms prior work on AudioCaps and achieves competitive performance on Clotho, indicating that bridging the modality gap enables effective use of LLM reasoning in audio captioning.

\section{Acknowledgement}
\vspace{-0.6em}
This work was supported by Institute of Information $\&$ communications Technology Planning $\&$ Evaluation (IITP) grant funded by the Korea government (MSIT) (RS-2025-02215122, Development and Demonstration of Lightweight AI Model for Smart Homes).
\vspace{-1.2em}
\ninept
\setstretch{0.96}
\setlength{\bibsep}{0.1ex}
\bibliographystyle{IEEEbib}
\bibliography{shortstrings,strings}

\end{document}